\documentclass[twocolumn,showpacs,preprintnumbers,amsmath,amssymb]{revtex4}
\usepackage{graphicx}
\usepackage{dcolumn}
\usepackage{bm}

\begin{document}

\title{Manuscript Title:\\Comment on "Layering transition in confined molecular thin films: Nucleation and growth"}

\author{Saroj Kumar Nandi}
\email{snandi@physics.iisc.ernet.in}
 \affiliation{Centre for Condensed Matter Theory, Department of Physics, Indian Institute of Science, Bangalore-560012, India.}

\date{\today}
\begin{abstract}
When fluid is confined between two molecularly smooth surfaces to a few molecular diameters, it shows a large enhancement of its viscosity. From experiments it seems clear that the fluid is squeezed out layer by layer. A simple solution of the Stokes equation for quasi-two-dimensional confined flow, with the assmption of layer-by-layer flow is found. The results presented here correct those in Phys. Rev. B, {\bf 50}, 5590 (1994), and show that both the kinematic viscosity of the confined fluid and the coefficient of surface drag can be obtained from the time dependence of the area squeezed out. Fitting our solution to the available experimental data gives the value of viscosity which is $\sim$7 orders of magnitude higher than that in the bulk. 
\end{abstract}

\pacs{68.35.-p, 68.45.-v, 68.15.+e}
      
\maketitle

The observed giant enhancement of viscosity dealing liquids under confinement has 
led to many experiments \cite{Hu, demirel, wei,alsten}, computer simulations
\cite{cui, cummings, scheidler}, and some theory
\cite{persson,zilberman,urbakh}. When confined to a spacing of few monolayers
between two solid surfaces, fluids start behaving very differently than in the
bulk, exhibiting a dramatic increase in
viscosity of six to seven orders of magnitude, and perhaps
an yield stress \cite{Hu, klein}.
 
The literature is conflicting on the question whether the observed slowing down
of the dynamics in the confined fluid stems from a phase transition
\cite{klein}. For fluid particles of regular shape, some interpretations show
the possibility of surface induced crystallization \cite{thompson}, others of a
glass transition \cite{schoen}, both at temperatures well above the respective
bulk transition temperature. It is somewhat unsatisfactory, if not strange,
that the same basic experiment is interpreted in both terms. One important
conclusion can be drawn from these that our theoretical understanding of the
phenomenon is not really complete.
 
In computer simulations, using surfaces with periodicity commensurate with the bulk crystalline phase of the confined fluid, the fluid
showed an ability to sustain an applied static stress \cite{thompson}. But in the
simulation of Gao et al \cite{gao} with incommensurate surfaces, this does not happen.
This shows the
importance of commensurability. But the experimental systems studied so
far involve confined fluids with a variety of molecular shapes and
architecture, and are generally incommensurate with the surfaces. Therefore 
commensurability is probably not relevant \cite{zhu}. Actually confinement is a means
to introduce geometrical constraints \cite{carolyn} and this slowing down of
the dynamics should have a more general origin rather than a mere
commensurability.

The high increase in the viscosity is perhaps a result 
of mode-coupling \cite{mct1,mct2} 
modified by confinement \cite{moumita}. Such a theory is beyond the scope of this 
work. It is however interesting that a simple phenomenological 
treatment in the form of layer-by-layer viscous flow 
gives physically relevant results. 
 
This model introduced here was first considered by Persson and
Tosatti\cite{persson}. But their solution does not say anything about the
kinematic viscosity, since their calculation {\em seemed} to show that the spreading
of the monolayers is determined by the wall-fluid friction, which they have
introduced as a separate parameter, and not by the kinematic viscosity of the
fluid. In fact the spreading {\em does} depend on the viscosity, as shown below. The
Persson and Tosatti model is thus more interesting than its authors showed it
to be. I redo the Persson and Tosatti calculation below with a result different
from theirs, and I point out where there was an error in their calculation. The
result is important because it shows that a measurement of the time evolution of
the area of the squeezed monolayer gives an estimate of the viscosity.
The value of viscosity extracted by fitting the time-dependence of the area 
to the form obtained here is consistent with those measured in independent 
experiments \cite{alsten}.  
The value of $\eta$ (the coefficient of
drag by the solid surface on the liquid film) is obtained to be of the 
same order of magnitude as that found by Persson and
Tosatti 
\begin{figure}
\includegraphics[width = 8.6cm]{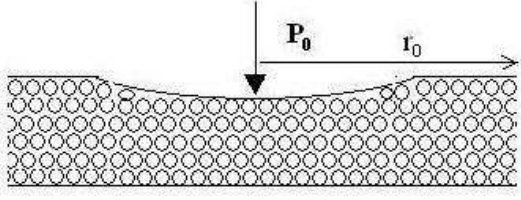} 
\caption{The suggested flow geometry
of Persson and Tosatti. The squeezing flow results in the $n$th layer becoming
an annulus with a central hole of radius $r$, which spreads and pushes the
rest of that layer outward, while the $(n-1)$ layers below are unaffected.}
\end{figure}

Let us take the geometry to be circular of radius $r_{0}$. The plates,
confining the fluid are taken to be smooth and devoid of any imperfections.  
Let $P_{0}$ be the normal force applied to squeeze out the fluid. Let us assume
that when the fluid is confined to a few molecular thickness, monomolecular
layer of liquid are squeezed out one at a time as a result of the applied
normal force as shown in Fig.1. This crucial assumption, the main phenomenological input 
in this approach, is supported by the experiment of Becker and
Mugele \cite{becker}. As in the experiment of \cite{Hu}, the
upper limit of change of liquid density $\rho$ due to shear is $4\%$. So we can
take the fluid to be incompressible in our model.

Suppose one monolayer is squeezed out, and let 
${\bf v}({\bf r},t)$ be the in-plane velocity field of this monolayer 
as a function of the two-dimensional in-plane coordinate ${\bf r}$ and time $t$. 
Then incompressibility and the  
Navier-Stokes equations become: 
\begin{equation} 
\label{incomp}
\nabla \cdot  {\bf{v(r}},t)=0 
\end{equation} 
\begin{equation}
\label{navstokes}
\frac{\partial {\bf{v}}}{\partial t} + {\bf v} . \nabla {\bf{v}} = - \frac
{1}{mn_a}\nabla p + \nu\bigtriangledown ^2 {\bf{v}} - \eta {\bf{v}}
\end{equation}
 where, $p$ is the two-dimensional pressure, $\nu$ is the
kinematic viscosity, $\eta$ is the coefficient of drag force by the sheets on
the fluid, $mn_a$ is the two dimensional mass density.

The flow is very slow, so we ignore inertia in Eq. (\ref{navstokes}). The suggested flow
geometry of Persson and Tosatti is that the squeezing flow results in the $n$th
layer becoming an annulous with a central hole of radius $r$, which spreads and
pushes the rest of that layer outward, while the $(n-1)$ layer below are
unaffected. Then \begin{equation} {\bf{v}} = v(r,t)\hat r \end{equation}
\begin{equation} p=p(r,t).  \end{equation}

Using Eq. (3) in Eq. (1) we get, \begin{equation}
 \frac {1}{r} \frac
{\partial}{\partial r}(rv)=0 
\end{equation} 
or,
\begin{equation} \frac{v}{r}+
\frac{\partial v}{\partial r} =0.  \end{equation}

Equation (2) becomes \begin{equation} \frac {\partial v}{\partial t}=-\frac
{1}{mn_a}\frac {\partial p}{\partial r}+\nu \frac {1}{r} \frac
{\partial}{\partial r}(r\frac {\partial v}{\partial r})- \eta v, \end{equation}
where symmetry implies only the radial component is non-zero.

At this point, \cite{persson} makes  
a mistake by making the
viscosity term, i.e. the second term in right hand side of Eq. (7), to be equal
to zero using Eq. (6). But this term should actually be equal to $\nu \frac
{v}{r^2}$. Using this value Eq. (7) becomes 
\begin{equation} 
\frac {\partial
v}{\partial t}=-\frac {1}{mn_a} \frac {\partial p}{\partial r}+\nu \frac
{v}{r^2}-\eta v 
\end{equation} 
From Eq. (6), we get 
\begin{equation} 
\dot
r=v=\frac {B(t)}{r} 
\end{equation} 
\begin{equation} \frac {\partial v}{\partial
t}=\frac {\dot B}{r}, \end{equation} where B(t) is an integration constant,
which depends only on t. Now using the above two results in Eq. (8) we get,
\begin{equation} \frac {\partial p}{\partial r}=-mn_a\frac {(\dot B + \eta
B)}{r}+mn_a\nu \frac {B}{r^3} \end{equation} In solving this equation, let us
take the boundary condition as $p$=$p_0$ at $r$=$r_0$. Then, \begin{equation}
p(r)-p_0=-mn_a(\dot B +\eta B) \ln(\frac {r}{r_0})-\frac {mn_a\nu B}{2}(\frac
{1}{r^2}-\frac {1}{r_0^2}) \end{equation} Now using Eq. (9) and Eq. (10), the
above equation can be written as \begin{equation} p(r)-p_0=-\frac
{mn_a\eta}{4}\frac {d}{dt}(r^2)\ln(\frac {r^2}{r_0^2})-\frac {mn_a\nu}{4}\frac
{d}{dt}(r^2)(\frac {1}{r^2}-\frac {1}{r_0^2}).  \end{equation}

Since $r$ varies very slowly, we have ignored the second order time derivative of
$r^2$.  Now let us take $\pi r^2=A$ and $\pi r_0^2=A_0$. Then, \begin{equation}
\frac {dA}{dt}\ln(\frac {A}{A_0})+\frac {\nu \pi}{\eta}\frac {dA}{dt}(\frac
{1}{A}- \frac {1}{A_0})= - \frac {4\pi (p-p_0)}{mn_a\eta} \end{equation}
Solving the above equation, assuming that p is independent of r, we get
\begin{equation} 
\frac {A}{A_0}\ln(\frac {A}{A_0})-\frac {A}{A_0}+\frac {\nu
\pi}{\eta A_0}\ln(A)-\frac {\nu \pi A}{\eta A_0^2}=-\frac {4\pi
(p-p_0)t}{mn_a\eta A_0}+c 
\end{equation} 
where $c$ is a constant of integration
which must be evaluated from the boundary condition. Let $t=t^*$ be the time at
which the whole layer is squeezed out. Then for $t=t^*$, $A$=$A_0$ and putting
this value in the above equation we will have, 
\begin{equation} 
(\frac {A}{A_0}
+\frac{\nu \pi}{\eta A_0})\ln(\frac{A}{A_0})+(1-\frac{A}{A_0})(1+\frac{\nu
\pi}{\eta A_0})=\frac {4\pi P_0 a}{mn_a\eta A_0}(t^*-t) 
\end{equation}

Now the adiabatic work required to squeeze out one monolayer of fluid will be
equal to $(p-p_0)A$. This adiabatic work must be equal to the change in free energy
due to the process. The total change in free energy is $(2\gamma _{sl} +V_{ll}
-2\gamma _{sl} +P_0 a)A-V_{ll} A$, where  $\gamma _{sl}$ is the solid-liquid
interfacial term, $V_{ll}$ is the van der Waals interaction term.  Equating
these two terms we get $(p-p_0)=P_0 a$.

Now Eq. (16) gives the relation between the area of the squeezed layer as a
function of time. From experiments $P_0,~ A_0,~ t^*$ and the three dimensional
density, all are known. So if we can take a time trace of the area of the
squeezed monolayer, then by fitting equation (16) to that data with $\eta$ and
$\nu$ as two fitting parameters, in principle, we can have a value of both
$\eta$ and $\nu$. So calculating $\eta$ and $\nu$ is a matter of having 
good data.
 
Now such good data is lacking in the literature. So, we took a  
time-sequence of
pictures of the squeezing out of one mono-layer of OMCTS in the paper by Becker
and Mugele \cite{becker}. We obtained the area $A$ of the squeezed layer as
well as the total area $A_0$ of the mica sheets by an area-calculating
software. Lacking an absolute estimate of the areas, we extracted 
$\frac{A}{A_0}$ from the pictures, thus obtaining 
the time-evolution of $\frac{A}{A_0}$.
We took $P_0$ as the maximum applied normal pressure $P_0$ $\sim 2\times 10^7
N/m^2$ and $A_0=7\times10^{-9} m^2$. These are the typical values of $P_0$ and
$A_0$ \cite{alsten,becker,gee} used in the experiments.
\begin{figure}
\label{Fit}
\includegraphics[width=8.6cm]{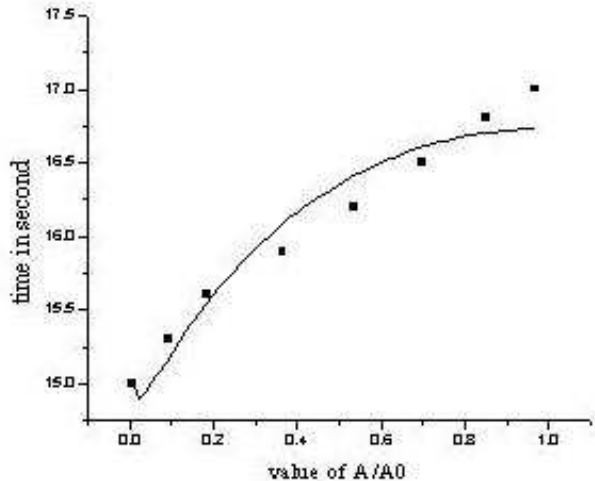}
\caption{The fit of Eq. (16) to the time sequence of squeezed area.
The data were obtained from a time sequence picture of the squeezed area in the paper of Becker and Mugele. The filled squares are the data points and the solid curve is the equation fit to the data.}
\end{figure}

With the values of the parameters quoted above we fitted Eq. (16) to the obtained data, shown in Fig.2. The fit
was not very good, since there are two parameters and only 8 data-points, 
but it was as much as could be done with the available information.  
The fit gave the values of $\eta \sim
7\times10^{14} s^{-1}$ and $\nu \sim 4\times10^5$ cm$^2$/s. The value of
$\eta$ is essentially same as that obtained by Persson and Tosatti. 
For comparison, the 
kinematic viscosity of OMCTS in bulk is $0.02$ cm$^2$/s.  
The value of $\nu$ from the fit 
is of the  same order of magnitude as that for confined OMCTS in the 
experiments of \cite{Hu,zhu}. The experiments of \cite{becker} thus 
indicate an enhancement of viscosity by about $\sim 7$ orders of magnitude. 
The above result is interesting in at least two senses.

(a) This is an approximate calculation and can not be expected to give
quantitatively accurate results. In this sense it is reassuring that 
a fit to our theory gives a value for $\nu$ very similar 
to that in independent experimental measurements.
Crude approximations of layer-by-layer flow and $p$ independent of $r$ 
are made to simplify the model, and the
dependence of viscosity on the film thickness can not be calculated from the
model. But at molecular thicknesses a liquid film supports a state of normal
stress and the film thickness adjusts itself to the externally applied normal
pressure. So, in this sense, the thickness and the normal pressure are not two
independent parameters.

(b)From Eq.(16), we see that one of the reasons of high increase of the value
of $\nu$ is the strong wall fluid interaction. The viscosity is directly related to the relaxation time of a fluid and in the simulation of Scheidler, Kob and Binder \cite{scheidler}, it is indeed found that the nature of the confining walls directly affects the dynamics of the fluid.

In comparision with the results obtained by Persson and Tosatti
\cite{persson}, we see that the value of $\eta$ is almost same as in their
calculation. Using the values of the different parameters, noted above, $\eta
\sim 10^{14}$. But the value of $\nu$ was not obtained by their (incorrect) 
calculation.
The model is more interesting than their calculation showed it to be, because
it tells us that $\nu$ can be extracted from the experiments 
as well. 

The theoretical approach presented here and, hence, the assumption of 
layer-by-layer flow, can be tested by comparison with 
experiments in which the area 
squeezed out is measured carefully as function of time.
In addition, different surface treatments should give different 
degrees of anchoring and hence different values of $\eta$, 
the coefficient of surface drag. The trends predicted by Eq. (16) 
with respect to changes in $\eta$ can then be compared with 
experiment.

The results presented here show that though the model is very simple and unable to
explain the phenomenon completely, it nonetheless demands at least a little
attention. 
\begin{acknowledgements}
Finally, I would like to express my gratitude to Professor Sriram Ramaswamy for
his helpful, important and enthusiastic discussions, suggestions and comments.
\end{acknowledgements}

\end{document}